\documentclass[a4paper]{ar-1col}
\usepackage{arydshln}
\usepackage[numbers]{natbib}
\usepackage{bm}
\usepackage{url}
\usepackage{amsmath,amssymb,amsthm, amsfonts, amsxtra}
\usepackage{pifont}
\usepackage{subfigure}
\usepackage{multirow}
\usepackage{bbold}
\usepackage{simplewick}

\newcommand{\bra}[1]{\< #1 \right|}
\newcommand{\ket}[1]{\left| #1 \>}
\definecolor{purple}{rgb}{0.5, 0, 0.5}
\renewcommand{\>}{\right\rangle}
\newcommand{\<}{\left\langle}   \newcommand{\bit}{\begin{itemize}}
\newcommand{\eit}{\end{itemize}}    

\newcommand{\be}{\begin{equation}}
\newcommand{\ee}{\end{equation}}
\newcommand{\ba}{\begin{align}}
\def\braket#1#2{\left\langle #1|#2\right\rangle}

\usepackage{xr}
\usepackage{amsmath}
\usepackage[ruled,vlined]{algorithm2e}
\usepackage[]{graphicx}
\usepackage{grffile}
\usepackage{mathrsfs}
\usepackage{framed}

 \usepackage{bbm}
 \usepackage{bm}
 \usepackage{braket}
 \usepackage{mathtools}

\usepackage[]{qcircuit}

\jname{Submitted to Annual Review of Condensed Matter Physics}
\jyear{2022}

\begin{document} 

\markboth{Chandran Iadecola Khemani Moessner}{Quantum Many Body Scars}

\title{Quantum Many-Body Scars: A Quasiparticle Perspective}

\author{Anushya Chandran$^1$, Thomas Iadecola$^{2,3}$, \\ Vedika Khemani$^4$,  Roderich Moessner$^5$
\affil{$^1$Department of Physics, Boston University, Boston, Massachusetts 02215, USA}
\affil{$^2$Department of Physics and Astronomy, Iowa State University, Ames, Iowa 50011, USA}
\affil{$^3$Ames Laboratory, Ames, Iowa 50011, USA}
\affil{$^4$Department of Physics, Stanford University, Stanford, CA 94305, USA}
\affil{$^5$Max-Planck-Institut f\"{u}r Physik komplexer Systeme, 01187 Dresden, Germany}}

\begin{abstract}

Weakly interacting quasiparticles play a central role in the low-energy description of many phases of quantum matter. 
At higher energies, however, quasiparticles cease to be well-defined in generic many-body systems due to a proliferation of decay channels. 
In this review, we discuss the phenomenon of quantum many-body scars, which can give rise to certain species of stable quasiparticles throughout the energy spectrum. 
This goes along with a set of unusual non-equilibrium phenomena including many-body revivals and non-thermal stationary states. 
We provide a pedagogical exposition of this physics via a simple yet comprehensive example, that of a spin-1 XY model. 
We place our discussion in the broader context of symmetry-based constructions of many-body scar states, projector embeddings, and Hilbert space fragmentation. 
We conclude with a summary of experimental progress and theoretical puzzles.

\end{abstract}

\maketitle

\tableofcontents

\section{Introduction}

Interacting quantum matter can exhibit  excitations  --- e.g.\ phonons in solids or magnons in magnets --- that involve the collective motion of many particles and yet, remarkably, act like a single emergent entity. Such objects are dubbed \emph{quasiparticles}.
The emergence of long-lived quasiparticles in very different types of condensed matter systems is so commonplace that one barely registers just how remarkable a phenomenon it is.  

There are various well-established mechanisms for stabilizing quasiparticles at low energies as encoded, for instance, in Goldstone's theorem for broken continuous symmetries, or in Ginzburg's kinematic argument for long-lived quasi-electrons and quasi-holes in Fermi liquids~\cite{cha95,coleman_2015,moessner_moore_2021}. 
In contrast, stabilizing quasiparticles at higher energies is a different issue altogether. It is commonly believed that quasiparticles become unstable when they encounter
the continuum of many-particle excited states at high energies, and decay swiftly when kinematically permitted. 

The existence of long-lived quasiparticles at energies well above the single-quasiparticle bandwidth should therefore register surprise. 
In fact, there exist various routes towards stabilizing  quasiparticles beyond the low-energy limit. Perhaps the most straightforward of these arises in the case of noninteracting particles, where long lifetimes arise due to the complete absence of decay processes. This comprises rather important phenomena such as Anderson localization as well as any noninteracting---e.g., topological---band structures, and includes spin systems related to such models via Jordan-Wigner transformations. More elaborate are ``interacting integrable" models, such as low-dimensional models soluble by the Bethe ansatz~\cite{Bethe_1931,Giamarchi03,Sutherland04,Essler05}, where particles scatter without decaying. The presence of certain symmetries can also guarantee the existence of infinitely long-lived quasiparticles, as in the $\eta$-pairing eigenstates of the Fermi-Hubbard model~\cite{Yang89}. Further, sufficiently strong interactions may actually impede, rather than promote, quasiparticle decay in certain energy windows~\cite{Bhatt:1974, Basko:2017,Verresen:2019}.

The subject of this review concerns a new mechanism of obtaining stable quasiparticles in interacting systems, which goes under the heading of ``quantum many-body scars." This term was coined in Ref.~\cite{Turner18a} by analogy to the phenomenon of wavefunction scarring in single-particle quantum mechanical systems arising from the quantization of a corresponding classical model~\cite{Heller84}. In such systems, quantum mechanical wavefunctions exhibit regions of enhanced probability, called scars, concentrated around unstable periodic trajectories of the classical system. The analogy posited in Ref.~\cite{Turner18a} defines quantum \emph{many-body} scars as the existence of special highly-excited eigenstates with atypically large weights on certain ``simple" many-body states. The dynamics of the many-body system, when prepared in one of these simple states, becomes strongly nonthermal. For example, the system can exhibit finite-time revivals of the {\it many-body} wavefunction, which generically occur only at astronomically long times (associated with Poincar\'e recurrence) in nonintegrable systems~\cite{CamposVenuti15}. While a variety of mechanisms for quantum many-body scars have been proposed, a ubiquitous one is the emergence of a small subset of stable quasiparticles, out of which special many-particle states throughout the many-body spectrum can be constructed~\cite{Moudgalya18a,Moudgalya18b,Schecter19,Iadecola20,Mark20a,Mark20,Moudgalya20,Chattopadhyay20,O'Dea20,Pakrouski20,Pakrouski21,Ren21,Tang21,Langlett22,Omiya22}. In idealized models, such quasiparticles have infinite lifetime, in contrast to the naive expectation outlined above.

These phenomena represent a deviation from the expected behavior of generic (i.e., interacting, nonintegrable) quantum many-body systems. The coherent quantum many-body dynamics of such systems has been the subject of a separate thread of investigation of much current interest. Nucleated by experimental developments---e.g., the advent of analog and digital quantum simulation platforms with low levels of environmental decoherence~\cite{Houck12,Langen14,Morgado20,Monroe21}---one important question has been the connection between microscopic quantum theory and macroscopic thermodynamics~\cite{D'Alessio16,Deutsch18}. The eigenstate thermalization hypothesis (ETH)~\cite{Jensen:1985aa,Deutsch91,Srednicki94,Rigol08} is a central concept in this context. It states that an eigenstate, $|\alpha\rangle$, of a many-body Hamiltonian, $H|\alpha\rangle=E_\alpha|\alpha\rangle$, exhibits the same properties as the appropriate Gibbs ensemble at the energy density $\epsilon_\alpha=E_\alpha/V$ to leading order in the volume $V$ of the system. For example, for any few-body operator $O$, the eigenstate expectation value $\langle\alpha|O|\alpha\rangle$ is a smooth function of $\epsilon_\alpha$. Similarly, if one partitions the system into two subregions $A$ and $B$ and calculates the reduced density matrix $\rho_A=\text{tr}_{B}\ket{\alpha}\bra\alpha$, the entanglement entropy
$S_A=-\text{tr}(\rho_A\ln\rho_A)$ scales extensively with the volume of $A$ at any finite temperature with the coefficient set by the thermal entropy density at the energy density $\epsilon_\alpha$.  In different words, the exponentially large amount of microscopic information encoded in the coefficients of the wavefunction on Hilbert space appears to be largely superfluous when determining thermodynamic observables. 

A many-body eigenstate in which expectation values of few-body observables differ from the thermodynamic values
is thus at odds with eigenstate thermalization. In the aforementioned examples of noninteracting and interacting integrable systems, the ETH as stated above does not apply and must be generalized to account for an extensive number of additional conserved quantities~\cite{Vidmar16}.
It is believed that the addition of a generic local perturbation of finite strength will destroy the special features associated with integrability and give way to conventional thermalizing behavior~\cite{Santos:2004vu,Rabson:2004nx,Santos:2010va,Modak:2014,Pandey:2020uo}. 
The one generic exception to the ETH is believed to be provided by many-body localization (MBL) in strongly disordered, yet interacting systems~\cite{Nandkishore15,Abanin19}, in which extensively many \emph{local} conserved degrees of freedom preclude thermalization~\cite{Serbyn13,Huse14,Swingle13}. 
The existence of such stable emergent integrability is however still being debated~\cite{Suntajs2020,Sels2021,Abanin:2021vq,Panda:2020aa,Crowley:2020,Morningstar:2021}.

Unlike systems with explicit or emergent integrability, systems with quantum many-body scars exhibit features that contrast with the ETH, but \emph{only} in a fraction of (excited) states that \textit{vanishes} in the thermodynamic limit. The vast majority of eigenstates instead look thermal. 
The non-thermal features may be broadly classified into three categories:
\begin{enumerate}
\item The existence of certain species of infinitely long-lived quasiparticles, 
\item Persistent many-body revivals at finite times in the dynamics, and
\item Non-thermal stationary states (e.g., eigenstates or late-time states under time evolution) that have atypically low (\emph{i.e.}, sub-thermal) entanglement and large weights on certain simple states.
\end{enumerate}
All of these items signal an absence of thermalization, touching on different aspects of the phenomenon. We reiterate that these non-thermal features are found in special states sprinkled \emph{throughout} the many-body spectrum --- \emph{i.e.}, at all temperatures, including infinite temperature --- and these special states coexist with typical thermal states at the same energy densities. The nomenclature of scarring has variously been used in the literature to signify the presence of any one of the above features, or of all three at the same time. In this review, we discuss quantum many-body scars from a quasiparticle perspective, and draw comparisons and contrasts with other perspectives on scarring and related phenomena.
Two other recent reviews on quantum many-body scars provide complementary perspectives to this one. 
The first, Ref.~\cite{SerbynReview:2021}, highlights  the connections to classical periodic orbits using semi-classics, while the second, Ref.~\cite{MoudgalyaReview:2021}, presents various exact results and matrix-product-state based understanding. 

The remainder of this review is organized as follows. For pedagogical simplicity and clarity, we introduce quantum many-body scars in Sec.~\ref{sec:S1XY} via a particularly simple model which exhibits the phenomena 1.--3.~in a physically transparent way. We then discuss quasiparticle-based many-body scarring more generally, focusing on a descriptive formalism in Sec.~\ref{sec:QPDescriptive} and constructive principles for building models with scars in Sec.~\ref{sec:QPConstructive}. 
In Sec.~\ref{sec:Zoology}, we explore other perspectives on scarring with the same emphasis on descriptive and constructive approaches. Finally, in Sec.~\ref{sec:Experiments}, we describe experimental observations of quantum many-body scars in various systems. We especially focus on the original experimental observation of scarring in a Rydberg system presented in Ref.~\cite{Bernien17}, and highlight theoretical attempts to explain this experiment via studies of the so-called ``PXP model" and its deformations; again, we focus on various manifestations of quasiparticles in these different accounts. 


\section{Quantum Scars from Stable Quasiparticles}

\subsection{An Illustrative Solvable Example}
\label{sec:S1XY}
In this section, we show how the three phenomena characteristic of quantum scarring emerge in a simple model, the spin-1 XY model. We focus on this model for pedagogical clarity, but note that the phenomenon of quasiparticles emerging at finite energy density in a nonintegrable model was first predicted in Refs.~\cite{Moudgalya18a,Moudgalya18b} for the Affleck-Kennedy-Lieb-Tasaki (AKLT) chain~\cite{Affleck87}.

\begin{figure}[t!]
\includegraphics[width=1.00\columnwidth]{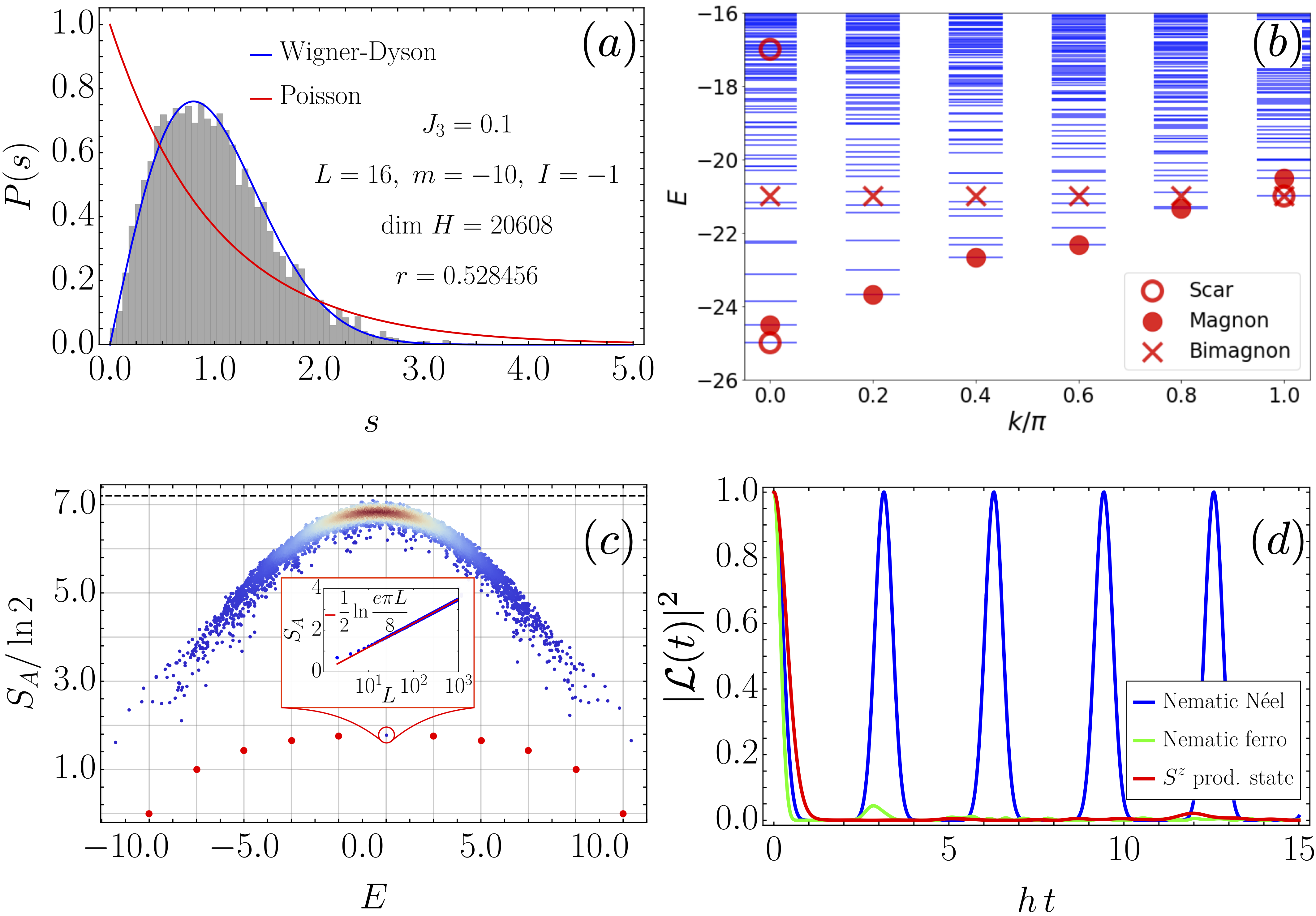}
\caption{
Scars in the spin-1 XY model (panels (a), (c), and (d) adapted from Ref.~\cite{Schecter19}). (a) Distribution $P(s)$ of nearest-neighbor energy level spacings $s$ for the model defined in Eq.~\eqref{eq:S1XY} in the presence of third-neighbor hopping terms (Eq.~\eqref{eq:H3}). The energy-level spacings follow a Wigner-Dyson distribution (blue curve), a strong indicator of nonintegrability. (b) Low-energy spectrum of the spin-1 XY model as a function of momentum $k$, with $L=10$ and parameters chosen such that the polarized state $\ket{\Omega}$ is the ground state. The scar states $\ket{\mathcal S_n}$ for $n=0,1,2$ are marked with empty red circles, while the single-magnon states described in Sec.~\ref{sec:NonSpecial} are marked with filled red circles. Red crosses denote the energy expectation values of single-bimagnon states for different values of $k$; only the one with $k=\pi$ is an exact eigenstate.  (c) Half-chain entanglement entropy $S_A$ as a function of energy in the zero-magnetization sector for a chain with $L=10$ sites. The scar state in the zero magnetization sector is circled in red, and the inset demonstrates that the entanglement entropy of this state scales logarithmically with $L$. The entanglement entropies of the scar states in other magnetization sectors are denoted by red points. The dashed line denotes the value of the entanglement entropy for a random state, which obeys volume-law scaling (see Ref.~\cite{Page93}). (d) Dynamics of the Loschmidt echo, Eq.~\eqref{eq:S1XYLt}, for the initial state Eq.~\eqref{eq:S1XYinit} (blue), with the dynamics for two other initial product states (green, red) shown for comparison.
}
\label{fig:S1XY}
\end{figure}

We follow the approach in Ref.~\cite{Schecter19}, and discuss a one-dimensional version of the model to simplify the presentation. However, many of the properties hold on the $d$-dimensional hypercubic lattice or indeed any bipartite graph. The model is given by
\begin{align}
\label{eq:S1XY}
\begin{split}
    H&=J\, H_{\rm XY}+h\, S^z\\
    &=J\sum_{r}(S^x_r S^x_{r+1} + S^y_r S^y_{r+1})+h\sum_r S^z_r,
\end{split}
\end{align}
where $r=1,\dots,L$ labels lattice sites. The operators $S^\alpha_r$ ($\alpha=x,y,z$) are spin-1 operators, and we choose a local basis $\ket{\pm_r}$, $\ket{0_r}$ whose eigenvalues under $S^z_r$ are $\pm 1$ and $0$, respectively. The model in Eq.~\eqref{eq:S1XY} is nonintegrable, even in one dimension, and therefore is expected to obey the ETH upon resolving all microscopic symmetries. This can be shown numerically by computing, \emph{e.g.}, the statistics of spacings between neighboring energy levels~\cite{Mehta04}, see Fig.~\ref{fig:S1XY}(a). Nevertheless, the model supports a tower of exact many-body eigenstates with a simple structure and atypical properties relative to other eigenstates at the same energy density. These scarred eigenstates have an appealing interpretation in terms of many-particle states of ``bimagnon" quasiparticles (\emph{i.e.}, bound states of two magnons), whose scattering and decay are inhibited by a destructive interference effect that is absent for other quasiparticle types, \emph{e.g.}~single magnons.

\subsubsection{Exact (Non-Scarred) Eigenstates Dictated by Symmetry}
\label{sec:NonSpecial}
The model in Eq.~\eqref{eq:S1XY} has two bands of exact single-magnon states in its spectrum due to a symmetry. To see this, note that the polarized states $\ket{\Omega}=\bigotimes^L_{r=1}\ket{-_r} $ and $\ket{\Omega'}=\bigotimes^L_{r=1}\ket{+_r}$ are annihilated by $H_{\rm XY}$. Single-magnon states can be built atop these polarized states using the spin raising and lowering operators $S^\pm_r=S^x_r\pm i\, S^y_r$:
\begin{subequations}
\label{eq:S1XYSingleMagnon}
\begin{align}
    \ket{+,k} &= \frac{1}{\sqrt{2L}}\sum^L_{r=1}e^{i k r} S^+_{r}\ket{\Omega}\\
    \ket{-,k} &= \frac{1}{\sqrt{2L}}\sum^L_{r=1}e^{i k r} S^-_{r}\ket{\Omega'},
\end{align}
\end{subequations}
where $k = \frac{2\pi}{L}p$ ($p=0,\dots,L-1$). These single-magnon states have energies $E_{\pm,k}=\mp h(L-1)+2J\, \cos k$. Their existence is guaranteed by the U(1) symmetry $[H,S^z]=0$, and by translation invariance. Indeed, the polarized states $\ket{\Omega}$ and $\ket{\Omega'}$ must be eigenstates of $H$ since they are the unique states with $S^z=\pm L$. 
There are $2L$ spin configurations with $S^z=\pm(L-1)$; translation invariance organizes these states into the single-magnon plane waves of Eqs.~\eqref{eq:S1XYSingleMagnon} Fig.~\ref{fig:S1XY}(b) plots the low-energy spectrum of Eq.~\eqref{eq:S1XY} with an added third-neighbor exchange term (see Eq.~\eqref{eq:H3}) as a function of quasi-momentum $k$, with parameters chosen such that the fully polarized state $\ket{\Omega}$ is the ground state. The exact single-magnon eigenstates $\ket{k,\Omega}$ with $k=0,\dots,\pi$ are represented with filled red circles.

Symmetry alone is not sufficient to guarantee the existence of eigenstates with more than one magnon. Indeed, one readily finds that inserting additional magnons via multiple application of the operators $S^{\pm}_{k} = \sum_r e^{ikr}S^{\pm}_{r}$ does not yield exact eigenstates. (This can be seen, \emph{e.g.}, by calculating the energy variance of these multimagnon states.) Physically, this arises because multi-magnon states are unstable to interaction effects (\emph{i.e.}, scattering processes) that open up various decay channels. Thus, other effects must conspire to stabilize eigenstates with additional quasiparticles.


\subsubsection{Bimagnon Tower of Scar States}

We now show that Eq.~\eqref{eq:S1XY} admits a tower of exact \textit{bimagnon} states,
\begin{align}
\label{eq:S1XYTower}
    \ket{\mathcal S_n}=\mathcal N(n) (J^+)^n\ket{\Omega},
\end{align}
where $n=0,\dots, L$, $\mathcal N(n)=\sqrt{\frac{(L-n)!}{n!L!}}$ ensures normalization, and where
\begin{align}
\label{eq:Jpmdef}
    J^\pm = \frac{1}{2}\sum^L_{r=1} e^{i\pi r} (S^\pm_{r})^2.
\end{align}
The operator $J^+$ inserts a bimagnon with momentum $\pi$ into the vacuum $\ket{\Omega}$.
The lowest-lying bimagnon state $\ket{\mathcal S_1}$ is marked with an empty red circle at $k=\pi$ in Fig.~\ref{fig:S1XY}(b), as are the polarized state $\ket{\mathcal S_0}=\ket{\Omega}$ and the state $\ket{\mathcal S_2}$. Single-bimagnon states at other momenta can also be constructed, and their energy expectation values are marked with red crosses in the same figure. However, these single-bimagnon states are not generically eigenstates for $k\neq \pi$.

The infinite lifetime of bimagnon excitations at momentum $\pi$ can be understood relatively simply. Consider the one-bimagnon state, which we rewrite as
\begin{align}
\label{eq:S_1}
    \ket{\mathcal S_1}\propto \sum^L_{r=1}(-1)^r\ket{\dots -+_r-\dots},
\end{align}
where we suppress the site indices away from site $r$ for compactness.
Observe that the local XY exchange term $h^{\rm XY}_{r,r+1}=S^x_rS^x_{r+1}+S^y_rS^y_{r+1}$ acts as
\begin{align}
\label{eq:hXY action}
\begin{split}
h^{\rm XY}_{r,r+1}\ket{--}&=h^{\rm XY}_{r,r+1}\ket{++}=0,\\
h^{\rm XY}_{r,r+1}\ket{+-}&=h^{\rm XY}_{r,r+1}\ket{-+}=\ket{00}.
\end{split}
\end{align}
The second line indicates that the XY exchange term dissociates the bimagnon excitation into two magnons.
From Eq.~\eqref{eq:hXY action}, we deduce that $h^{\rm XY}_{r,r+1}$ annihilates $\ket{\mathcal S_1}$ for all $r$:
\begin{align}
\label{eq:interference}
\begin{split}
    h^{\rm XY}_{r,r+1}\ket{\mathcal S_1}&\propto h^{\rm XY}_{r,r+1}\ket{\dots+_r-\dots}-h^{\rm XY}_{r,r+1}\ket{\dots-+_{r+1}\dots}=0.
\end{split}
\end{align}
This annihilation results from the destructive interference of the bimagnon dissociation processes $\ket{+-}\to \ket{00}$ and $\ket{-+}\to \ket{00}$, which arises due to the alternating sign $e^{i\pi r}=(-1)^r$ in Eq.~\eqref{eq:S_1} In fact, the above calculation can be applied to show that $h^{\rm XY}_{r,r+1}\ket{\mathcal S_n}=0$ for all $n$. Every spin configuration $\ket{L}\otimes \ket{+-} \otimes \ket{R}$ in $\ket{\mathcal S_n}$ has a partner configuration $\ket{L}\otimes \ket{-+} \otimes \ket{R}$ appearing with the opposite sign due to the fact that all of the bimagnons have momentum $\pi$. We therefore conclude that $H_{\rm XY}\ket{\mathcal S_n}=0$, so that $\ket{\mathcal S_n}$ are eigenstates of Eq.~\eqref{eq:S1XY} with energy $E_n=h(2n-L)$ set by the magnetic field.

An important feature of the states $\ket{\mathcal S_n}$ is that they form a spin-$L/2$ representation of an SU(2) algebra generated by the raising and lowering operators $J^\pm$:
\begin{align}
\label{eq:S1XYSU2Alg}
    [J^+,J^-]=2J^z,\indent [J^z,J^\pm] = \pm J^\pm,
\end{align}
where $J^z=\frac{1}{2}\sum^L_{r=1}S^z_r=\frac{1}{2}S^z$. Note that the states $\ket{\mathcal S_n}$ are eigenstates of $J^z$, which is proportional to the magnetization operator $S^z$ that couples to the magnetic field. Using the definition Eq.~\eqref{eq:S1XYTower} of the scar states, one can show that
\begin{subequations}
\label{eq:S1XYSU2Rep}
\begin{align}
    \left[\frac{1}{2}(J^+J^-+\text{H.c.})+(J^z)^2\right]\ket{\mathcal S_n}=j(j+1)\ket{\mathcal S_n}
\end{align}
and
\begin{align}
    J^\pm\ket{\mathcal S_n}=\sqrt{j(j+1)-m(m\pm 1)}\ket{\mathcal S_{n\pm 1}},
\end{align}
\end{subequations}
where $j=L/2$ and $m = n-L/2$.
Eqs.~\eqref{eq:S1XYSU2Rep} enable simple calculations of the expectation values of certain operators.

As might be anticipated from their simple structure (see Eq.~\eqref{eq:S1XYTower}) and physical interpretation, the eigenstates $\ket{\mathcal S_n}$ exhibit properties that are atypical of eigenstates nearby in energy. In particular, we should compare an eigenstate $\ket{\mathcal S_n}$ with $n/L$ finite as $L\to \infty$ to expectations based on the ETH, which governs typical eigenstates at the same energy density. One property that clearly demonstrates that the ETH does not hold in these eigenstates is the presence of off-diagonal long-range order~\cite{Yang62}, as witnessed by the correlation function
\begin{align}
\label{eq:J+J-}
    \frac{4}{L^2}\braket{\mathcal S_n|J^-J^+|\mathcal S_n} = \left[1-\frac{(2n-L)^2}{L^2}\right]+O\left(\frac{1}{L}\right).
\end{align}
This result is obtained by direct application of Eqs.~\eqref{eq:S1XYSU2Rep}
Notice that, unless $n=0$ or $L$, the right-hand side of the above equation is finite in the thermodynamic limit. This means that the scarred eigenstates $\ket{\mathcal S_n}$ contain long-range \textit{connected} correlations (note that $\braket{\mathcal S_n|J^\pm|\mathcal S_n}=0$). In contrast, for a generic eigenstate in the middle of the energy spectrum, the expected value of this correlation function based on ETH is given by the infinite-temperature average
\begin{align}
    \frac{1}{3^L}\text{tr}\left(\frac{4}{L^2}J^{-}J^{+}\right)=\frac{4}{3L},
\end{align}
which vanishes as $L\to\infty$. 
Another atypical feature of the eigenstates $\ket{\mathcal S_n}$ is their entanglement entropy, which can be evaluated analytically using simple combinatorics (see Ref.~\cite{Schecter19}). The scaling with $L$ of the entanglement entropy of the state $\ket{S_{L/2}}$ as $L\to\infty$ can be shown explicitly via saddle point arguments to be
\begin{align}
    \lim_{L\to\infty}S_{A}=\frac{1}{2}\left(\ln\frac{\pi L}{8}+1\right),
\end{align}
where $S_{A}=-\text{tr}(\rho_A\ln\rho_A)$, where $\rho_A =\text{tr}_B\ket{\mathcal S_{L/2}}\bra{\mathcal S_{L/2}}$ and $\text{tr}_B$ denotes the trace over sites $1,\dots,L/2$. In contrast, the ETH predicts that finite-energy-density eigenstates of nonintegrable models like Eq.~\eqref{eq:S1XY} obey the volume-law scaling $S_A\sim L$. Numerical results emphasizing the weakly-entangled nature of the scar states are shown in Fig.~\ref{fig:S1XY}(c).

Further evidence of the atypical nature of these eigenstates can be revealed by considering the dynamics from certain special initial states. In the presence of a finite magnetic field $h$, any state of the form
\begin{align}
\label{eq:Initial state}
    \ket{\psi_0}=\sum^{L}_{n=0} c_n\ket{\mathcal S_n}
\end{align}
experiences periodic revivals due to the equally spaced eigenenergies $E_n$. This is easiest to see in the so-called Loschmidt echo,
\begin{align}
\label{eq:S1XYLt}
\mathcal L(t) &= \braket{\psi_0|e^{-iHt}|\psi_0}=e^{i\, hLt}\sum^L_{n=0}|c_n|^2 e^{-i\, n(2h)t}.
\end{align}
Up to an unimportant global phase, $\mathcal L(t)$ is a periodic function with period $\pi/h$. Numerical data exhibiting these periodic revivals are shown in Fig.~\ref{fig:S1XY}(d).
This time-periodic behavior of the Loschmidt echo should be contrasted with the expectation for quantum chaotic systems, where a rapid decay is expected for typical initial states \cite{Gorin06,Goussev16}.
Remarkably, the family of initial states in Eq.~\eqref{eq:Initial state} includes product states that are readily accessible experimentally. Perhaps the simplest of these is given by
\begin{align}
\label{eq:S1XYinit}
    \ket{\phi_0}&=\bigotimes_r \left[\frac{\ket{+_r}-(-1)^r\ket{-_r}}{\sqrt{2}}\right],
\end{align}
which is of the form Eq.~\eqref{eq:Initial state} with $c_n=(-1)^n\sqrt{\binom{L}{n}\frac{1}{2^L}}$. Physically, the state $\ket{\phi_0}$ corresponds to the ground state of the Hamiltonian \begin{align}
\begin{split}
    H_0&=\frac{1}{2}(J^++J^-)\\
    &=\frac{1}{2} \sum^L_{r=1}(-1)^r\left[(S^x_r)^2-(S^y_r)^2\right]
\end{split}
\end{align}
formed from the quasiparticle creation and annihilation operators. $H_0$ is nothing but the generator $J^x$ of the SU(2) algebra Eq.~\eqref{eq:S1XYSU2Alg} As such, the periodic dynamics observed in Eq.~\eqref{eq:S1XYLt} can be interpreted as resulting from the precession of a macroscopic spin-$L/2$ object under the magnetic field term in Eq.~\eqref{eq:S1XY}, which is proportional to $J^z$. This interpretation was put forward as an approximate description of the periodic dynamics in the PXP model in Ref.~\cite{Choi19} (see Sec.~\ref{sec:Experiments}), but holds exactly in the case of the spin-1 XY model.

We conclude this section with a few comments regarding the model in Eq.~\eqref{eq:S1XY} First, we note that, in one dimension, the spin-1 XY model possesses a hidden nonlocal SU(2) symmetry when defined with open boundary conditions, and a ``twisted" version thereof when defined with periodic boundary conditions ~\cite{Kitazawa03,Chattopadhyay20}. Thus, confirming the nonintegrability of this model in one dimension requires resolving this additional unexpected symmetry. However, this symmetry is not essential to the presence of the scars in this model. For example, as noted in \cite{Schecter19}, adding a third-neighbor XY exchange term of the form
\begin{align}
\label{eq:H3}
    H_3=J_3\sum_{r}(S^x_r S^x_{r+3} + S^y_r S^y_{r+3})
\end{align}
breaks the hidden SU(2) symmetry while preserving the scar states. Indeed, one readily verifies that the destructive interference effect demonstrated in Eq.~\eqref{eq:interference} holds also for $H_3$, the key commonality being that sites $r$ and $r+3$ belong to opposite sublattices, as do sites $r$ and $r+1$. In fact, this reasoning can be extended to show that any odd-neighbor XY exchange term preserves the scar tower. A variety of other terms can also be added to Eq.~\eqref{eq:S1XY} while preserving the tower of scar states. For example, in \cite{Schecter19} it was noted that the single-site anisotropy term $\sum_r (S^z_r)^2$ preserves the tower, since the states $\ket{\mathcal S_n}$ are superpositions of tensor products of the local states $\ket{\pm_r}$, which satisfy $(S^z_r)^2\ket{\pm_r}=\ket{\pm_r}$. Ref.~\cite{Mark20} presented a systematic enumeration of all nearest-neighbor terms that preserve the eigenstates $\ket{S_n}$.

\subsection{Algebraic Descriptions} 
\label{sec:QPDescriptive}
At a more abstract level, models with stable quasiparticles atop specific vacuum states have simple algebraic descriptions.

There are three equivalent algebraic descriptions in the literature: spectrum generating algebras (SGAs)~\cite{Mark20a, Moudgalya20}, quasi-symmetry groups~\cite{Ren21, Ren22}, and group-invariant sectors~\cite{Pakrouski20,Pakrouski21}. All three describe the scarred subspace using higher symmetries than those of the Hamiltonian itself, but provide different constructive principles for the Hamiltonian and for the action of the symmetry on the scarred subspace (see Sec.~\ref{sec:QPConstructive}). The scar towers in the spin-1 XY chain discussed above, generalized Hubbard models with $\eta$-pairing states~\cite{Yang89,Vafek17,Mark20,Moudgalya20}, the AKLT spin-1 chain~\cite{Moudgalya18a}, transverse field Ising ladders~\cite{Voorden:2020wo}, a domain wall conserving spin-1/2 model~\cite{Iadecola20}, models with multiple magnons~\cite{Tang21, Mark20a}, and various chiral and non-chiral fermionic models~\cite{Martin:2022uj,Pakrouski20,Schindler:2022} can be understood using at least one of the three descriptions. As the SGA is most closely connected to the quasiparticle perspective, we describe it below.

A quasiparticle creation operator $Q^+$ provides an SGA if its commutator with the Hamiltonian $H_\mathrm{SG}$ satisfies:
\begin{align}
\label{Eq:SGAAll}
    [H_\mathrm{SG}, Q^+]  &= \omega Q^+ 
\end{align}
for $\omega \neq 0$~\footnote{Eq.~\eqref{Eq:SGAAll} has also been referred to in the literature as a dynamical symmetry~\cite{Buca19,Buca20,Buca22}.}. If $\ket{\psi_0}$ is an eigenstate of $H_\mathrm{SG}$ with eigenvalue $E_0$, then $(Q^+)^n \ket{\psi_0}$ is also an eigenstate with eigenvalue $E_0 + n \omega$ for any integer $n>0$ (as long as $(Q^+)^n \ket{\psi_0} \neq 0$). The SGA thus provides multiple towers of equally spaced eigenstates.

Often however, Eq.~\eqref{Eq:SGAAll} asks for too much and results in special towers of states for symmetry reasons. For example, consider the SU(2) algebra defined in Eq.~\eqref{eq:S1XYSU2Alg} that arises in the description of the scar tower in the spin-1 XY model. Suppose that we define a Hamiltonian of the form $H_{\rm SG}=H_\mathrm{sym} + \omega J^z$, where $H_\mathrm{sym}$ is $J$-SU(2) symmetric, i.e., it commutes with all of the generators in Eq.~\eqref{eq:S1XYSU2Alg} (An example of such an $H_{\rm sym}$ was found in Ref.~\cite{Mark20a}; see Eq.~\eqref{eq:S1XYHsym} below.) The $J$-SU(2) raising operator $J^+$ then defines an SGA as in Eq.~\eqref{Eq:SGAAll}, with $Q^+ = J^+$. In this scenario, the tower $\ket{\mathcal{S}_n}$ in Eq.~\eqref{eq:S1XYTower} is \textit{not} quantum many-body scarred, as each $\ket{\mathcal{S}_n}$ has maximal $J^2$ eigenvalue and is the only state in its symmetry sector. It is precisely because the Hamiltonian of the spin-1 XY Hamiltonian $H_{\rm XY}$ in Eq.~\eqref{eq:S1XY} is not $J$-SU(2) symmetric that the tower of states $\ket{\mathcal{S}_n}$ qualify as bona fide many-body scar states. 

It is therefore useful to define the notion of a \textit{restricted} SGA, which only requires that Eq.~\eqref{Eq:SGAAll} holds within the scarred subspace. That is, for a given vacuum state $\ket{\psi_0}$, if,
\begin{align}
\label{Eq:SGADefine}
\begin{split}
    H \ket{\psi_0} &= E_0 \ket{\psi_0}, \\
    [H, Q^+] \ket{\psi_n} &= \omega Q^+ \ket{\psi_n} \ \forall\ n\ \textrm{s.t.}\ (Q^+)^n \ket{\psi_0} \neq 0
\end{split}
\end{align}
then the states with $n$ quasiparticles atop the vacuum state, $\ket{\psi_n} = (Q^+)^n \ket{\psi_0}$, are exact eigenstates of a Hamiltonian $H$ with energies $E_0 + n \omega$. In practice, additional terms are added to SGA Hamiltonians to deform them into Hamiltonians that only satisfy Eq.~\eqref{Eq:SGADefine} Eq.~\eqref{eq:S1XY} represents one such deformation for the spin-1 case.

It is often the case that the stationarity of few quasi-particle states, $\ket{\psi_m}$ for $m\leq M$, guarantees the stationarity of states with any number of quasiparticles atop the vacuum state, $\ket{\psi_n}$. For example, in the spin-1 XY chain, $M=1$, while in the AKLT chain, $M=2$~\cite{Moudgalya18a}. A rewriting of the condition in Eq.~\eqref{Eq:SGADefine} in terms of nested commutators then becomes useful~\cite{Moudgalya20}:
\begin{subequations}
\label{Eq:nestedcommRSGA}
\begin{align}
    H \ket{\psi_0} &= E_0 \ket{\psi_0} \\
    \mathrm{ad}_{Q^+}(H) \ket{\psi_0 } &= -\omega Q^+ \ket{\psi_0 }\\
    \mathrm{ad}^m_{Q^+}(H) \ket{\psi_0 } &= 0, \,  2 \leq m \leq M \\
    \mathrm{ad}^m_{Q^+}(H) &= 0, \,\, \forall\ m> M.
\end{align}
\end{subequations}
Above $\mathrm{ad}_X(Y) = [X,Y]$. Solving the operator equation for $m>M$ is much simpler than solving the restricted equation for $2\leq m \leq M$, and provides valuable clues about the scarred states and/or the Hamiltonian $H$. For example, Ref.~\cite{Tang21} constructed operator bases that satisfied Eq.~\ref{Eq:nestedcommRSGA}d. for a given $Q^+$ and $M$, while Ref.~\cite{Moudgalya18a} identified bimagnons as candidate stable quasiparticles for AKLT spin-1 chains by solving Eq.~\ref{Eq:nestedcommRSGA}d. for $M=2$.



\subsection{Constructive Principles}
\label{sec:QPConstructive}
There are several constructive principles that provide restricted SGAs, and thus scarred Hamiltonians with certain species of stable quasiparticles. We focus here on a principle that leverages non-Abelian symmetries, and also discuss another mechanism based on particle confinement.

Non-Abelian symmetries $G$ furnish raising and lowering operators $Q^\pm$ that naturally satisfy an SGA with (Cartan) generators $Q^z$ in any subspace.
When coupled with $G$-symmetric Hamiltonians and symmetric base states, we can construct large families of Hamiltonians with quasiparticle scar states~\cite{O'Dea20}:
\begin{align}
\label{eq:Hsymmetryformalism}
    H = H_\mathrm{sym} + H_\mathrm{SG} + H_A. 
\end{align}
We now describe each of these terms.

The first term $H_\mathrm{sym}$ is $G$-symmetric, that is, $[H_\mathrm{sym}, Q^\pm]=0$ and $[H_\mathrm{sym}, Q^z]=0$. (Note that, in general, $H_{\rm sym}=0$ is also allowed.) Suppose the base state is an eigenstate of $H_\mathrm{sym}$. Then, $\ket{\psi_0}$ is labelled by its eigenvalue under the Casimir operator $Q^2$ and the Cartan generator $Q^z$.  Furthermore, $\ket{\psi_n} $ is a degenerate eigenstate with the same value of $Q^2$ but a different value of $Q^z$. To guarantee that the base state is an eigenstate of $H_\mathrm{sym}$, it is often chosen to be an element of a unique irreducible representation (irrep) of the group $G$. For example, in the spin-1 XY model, the scarred states span the unique maximum-spin representation of the SU(2) algebra in Eq.~\eqref{eq:S1XYSU2Alg}, to which the polarized state $\ket{\Omega}$ belongs. The $J$-SU(2) symmetric part of the spin-1 XY exchange Hamiltonian $H_\mathrm{XY}$ was found in Ref.~\cite{Mark20} to be given by
\begin{align}
\label{eq:S1XYHsym}
\begin{split}
H_{\rm sym} &= \sum_{r}\left(\ket{+_r0_{r+1}}\bra{0_{r}+_{r+1}}\ -\ \ket{-_r0_{r+1}}\bra{0_{r}-_{r+1}}\ +\ \text{H.c.}\right).   
\end{split}
\end{align}

The second term, $H_\mathrm{SG}$, is a (linear combination of) generator(s) in the Cartan sub-algebra of $G$ so that $[ H_\mathrm{SG},Q^{\pm}] = \pm \omega Q^\pm$.  This term lifts the degenerate multiplets of $H_\mathrm{sym}$ into energetically equidistant towers of states. In the spin-1 XY model, $H_\mathrm{SG} = h S^z=2h J^z$ so that $\omega = 2h$.

The final term in Eq.~\eqref{eq:Hsymmetryformalism} breaks the $G$-symmetry, but annihilates the scar states: $H_A \ket{\psi_n} = 0$. In the spin-1 XY model Eq.~\eqref{eq:S1XY}, it is given by $H_A = J(H_\mathrm{XY}-H_{\rm sym})$, where $H_{\rm sym}$ is given in Eq.~\eqref{eq:S1XYHsym} Explicitly~\cite{Mark20a},
\begin{align}
\begin{split}
    H_A &= J\sum_{r}\left[(\ket{+_r-_{r+1}}+\ket{-_r+_{r+1}})\bra{0_r0_{r+1}}\ +\ 2\ket{-_r0_{r+1}}\bra{0_r-_{r+1}}\ +\ \text{H.c.} \right].
\end{split}
\end{align}

The symmetry-based formalism is powerful because of its generality. First, $G$ can be any continuous non-Abelian symmetry group, or a $q$-deformed version thereof~\cite{O'Dea20,Ren21}. Indeed, there are examples of scarred models in the literature in which $G$ is SU(3)~\cite{O'Dea20}, $q$-deformed SU(2)~\cite{O'Dea20}, U($N$) or SO($N$) where $N$ is the number of lattice sites~\cite{Pakrouski20}, etc. Next, even when $G$ is SU(2), it need not be the spin-SU(2). Prominent examples include the Hirsch model with $\eta$-pairing states as scarred states~\cite{Mark20} and generalized AKLT models~\cite{Moudgalya20b}. Third, the scar tower need not be an irrep of $G$ with maximal eigenvalue under the Casimir operators. In the SU(2) case, for example, the base state could be a single-magnon state with zero quasi-momentum and $S = L/2-1$~\cite{Tang21}. Indeed, the scar tower need not be an irrep of $G$ at all if $H_\mathrm{sym}$ is special and multiple irreps are degenerate. The scarred bimagnon tower in the AKLT model provides such an example~\cite{Mark20a}. Fourth, the same symmetry group $G$ can provide multiple raising operators $J^+$ to embed scar pyramids~\cite{Tang21,Mark20a}. For example, with $G=$~SU(2), single magnon excitations with multiple momenta ($k=0$ and $k=k_0$ for any $k_0$) can be infinitely long-lived excitations atop the fully polarized state. 

There are a few examples of scarred models that do not obviously fit into the symmetry-based formalism. For instance, in the domain-wall conserving model of Ref.~\cite{Iadecola20} the quasiparticle creation operator $Q^+$ does not a have a clear connection to a root of a Lie algebra. Another example, surprisingly, arises in the spin-1 XY model, which hosts a second tower of so-called ``bond-bimagnon" states~\cite{Schecter19, Chattopadhyay20}. The quasiparticle creation operator for this tower of states is so far not known. However, in both of these examples, the towers of states arise from the projection of an auxiliary tower of states that \textit{do} transform as a representation of a Lie algebra~\cite{Iadecola20,Chattopadhyay20,Ren22}.

An interesting application of the symmetry-based formalism is to deformations of integrable points. Integrable points have extensively many continuous symmetries, and thus provide extensively many creation operators $Q^+$ of distinct quasiparticles. A suitable deformation $H_A$ could preserve one or several of these quasiparticles as infinitely long-lived excitations. The deformation of integrable points thus provides a systematic way to construct large families of quantum scarred models. An early study of a deformed PXP model suggested that this model is proximate to integrability~\cite{Khemani19}, which might explain the origin of scarring in it (see Sec.~\ref{sec:Experiments} for further discussion). Integrability also played an important role in scars that were experimentally obtained in a 1D dysprosium quantum gas~\cite{Wil:2021vr}. Theoretically, the connection between scars and integrability has been studied in Refs.~\cite{Moudgalya20b, Shibata:2020,Medenjak:2020,O'Dea20,Martin:2022uj,Zhang22b}, but this is still largely unexplored territory. 

How to construct the all-important $H_A$ that breaks the global symmetry? There are several methods discussed in the literature. 

One method is to use local projectors that annihilate local patterns that appear in the scarred tower. Such projectors can be constructed using the symmetry $G$ itself. For example, Refs.~\cite{Ren21,Ren22} construct the subspace generated by the action of $G$ on the base state restricted to a small number of sites; the complement of this subspace provides the required local projector. The result is a frustration-free Hamiltonian, discussed further in Sec.~\ref{Sec:SMEmbed}. 

Another method is to choose the group $G$ so that the scarred subspace is a singlet of the group~\cite{Pakrouski20,Pakrouski21}. In this case, any $H_A = \sum_a O_a T_a$, where $T_a$ are the generators and $O_a$ are any operators that make $H_A$ Hermitian, annihilate the scarred subspace. 
Note that, in this case, $H_\mathrm{sym}$ need not be $G$-symmetric---it need only have the $G$-singlets as eigenstates, see Refs.~\cite{Pakrouski20,Pakrouski21} for more details.

Brute force computation in an operator basis is also useful. That is, we search for coefficients $c_a$ of a set of translationally invariant operators $O^a = \sum_i O^a_{i}$ such that $H_A = \sum_a c_a O^a$ annihilates all the states in the scarred manifold. When the SGA is furnished by a symmetry, choosing the operators $O^a$ to be irreducible tensor representations sets several $c_a$ to zero on symmetry grounds~\cite{Tang21}. The remaining $c_a$ satisfy simple linear relations, which can be solved analytically or numerically~\cite{Tang21,Qi2019determininglocal, O'Dea20}. The resulting $H_A$ generically does not annihilate the scarred states term-by-term. Such Hamiltonians are thus referred to as ``as-a-sum" annihilators.

Other methods use the language of matrix product states and operators to construct nearest neighbor Hamiltonians with a quasiparticle tower of scars~\cite{Moudgalya20b,Ren22}. The starting point is to write the states with $n$ quasiparticles, $\ket{\psi_n}=(Q^+)^n \ket{\psi_0}$, as matrix product states using two types of matrices for each site depending on whether there is or is not a quasiparticle on that site. Enforcing that the quasiparticles cannot sit on neighboring sites results in simple algebraic conditions on parent Hamiltonians with the scarred tower $\ket{\psi_n}$. This method was used to discover several AKLT-type models with bimagnon scars atop a spin liquid ground state~\cite{Moudgalya20b,Mark20a}, perturbed Potts models with tri-magnon scars~\cite{Moudgalya20b}, and a two-dimensional model with topologically ordered states within the scarred subspace~\cite{Ren22}.

As an aside, we note that none of the constructions described above guarantee that $H_A$ exclusively annihilates the scarred subspace. Indeed, determining the number of states annihilated by a local Hamiltonian is in general intractable, although interesting lower bounds exist~\cite{Sattath:2016}. 

Refs.~\cite{James:2019,Robinson:2019} suggest that particle confinement leads to quantum many-body scars at varying energy densities. Specifically, Refs.~\cite{James:2019,Robinson:2019} study the ordered phase of the Ising model in transverse and longitudinal fields in one dimension. The longitudinal field leads to confinement of domain walls, so that a pair of domain walls forms a bound state (a meson) whose energy grows with their separation. These single meson states persist as eigenstates well above the two-meson threshold; they apparently persist even when the two domain walls are macroscopically separated and the state has a finite energy density above the ground state. Whether the persistence is only a finite-size effect in the lattice model, and whether other continuum models that are in confined phases exhibit scars, are open questions.  

\section{Generalized Scars Beyond Quasiparticles}
\label{sec:Zoology}

In this section, we broaden our scope beyond quasiparticles and return to the connection of scars with the failure of thermalization mentioned in the introduction. The purpose of this section is, in part, to chart out the historical genesis of scars, and also to provide an overview of the richness of phenomena that have been associated with term scar, but which at times refer to some rather distinct types of underlying physics. 

\subsection{Scars from Projector Embeddings}
\label{Sec:SMEmbed}
Perhaps the earliest mention of the existence of a special eigenstate of a Hamiltonian that passes robustly, with perfectly unavoided crossings, through other states as parameters of the Hamiltonian are varied was provided by Shastry and Sutherland~\cite{Shastry:1981}. In their eponymous spin model, an exactly dimerized state remains an eigenstate even when it ceases to be the ground state of the model.  This kind of scar state does not come with a tower of states at all---it is \emph{a priori} just an individual very special state which happens to escape eigenstate thermalization.

The Shastry-Sutherland model is an example of an important class of Hamiltonians known as frustration-free Hamiltonians. Such Hamiltonians can be written (up to a constant energy shift) as a sum of local projection operators $P_i$,
\begin{align}
\label{eq:HFF}
H_{\rm FF}=\sum_i h_i P_i,
\end{align}
with real coefficients $h_i\geq 0$. The ground state(s) of frustration-free Hamiltonians are states $\ket{\psi_{\rm FF}}$ that satisfy $P_i\ket{\psi_{\rm FF}}=0$ for all $i$. Note that two projectors $P_i$ and $P_j$ in the Hamiltonian Eq.~\eqref{eq:HFF} need not commute with each other, so that $H_{\rm FF}$ may be nonintegrable. In addition to the Shastry-Sutherland model, well-known examples include the AKLT and Klein models~\cite{Affleck87,Chayes_1989}, the Rokhsar-Kivelson model~\cite{Rokhsar88} and its generalizations~\cite{Castelnovo05}, the toric code~\cite{Kitaev:2003vn}, and other parent~\cite{Fannes92,Nachtergaele96,PerezGarcia07} and uncle~\cite{Fernandez15} Hamiltonians for matrix product states.

It is always possible to modify a frustration-free Hamiltonian such that a ground state of the original model becomes an arbitrarily highly excited state of the modified model. For example, in Eq.~\eqref{eq:HFF}, one can modify the coefficients $h_i$ to have both positive and negative signs, in addition to varying magnitudes, or even promoting $h_i$ to arbitrary local operators. The erstwhile ground states $\ket{\psi_{\rm FF}}$ then remain zero-energy states of the modified Hamiltonian. 

The connection of these ideas to the ETH was first drawn by Shiraishi and Mori (SM) in Refs.~\cite{Shiraishi17,Mori17a}. The SM construction again relies on the existence of an extensive set of local projectors $\{P_i\}$, and a target scarred subspace $\mathcal{T}$ spanned by states annihilated by the local projectors: $P_i|\psi\rangle =0\; \forall \; i$. The SM Hamiltonian is a sum of two terms: the first is a (modified) frustration-free Hamiltonian $H_A$ that annihilates any state in the target subspace $\mathcal{T}$, and the second term $H'$ commutes with all the projectors $P_i$ ($[H', P_i]=0$):
\begin{equation}
    H_{\rm SM} = \underbrace{\sum_i P_i h_i P_i}_{H_A} + H',  
\end{equation}
where $h_i$ are arbitrary local Hamiltonians. 
The Hamiltonian is constructed to be block-diagonal since $H'$ does not mix states in $\mathcal{T}$ with those in its complement $\overline{\mathcal{T}}$. Thus, $H_{\rm SM}$ can be diagonalized in $\mathcal{T}$ and $\overline{\mathcal T}$ independently. 
The energy eigenvalues of the eigenstates $\{\ket{\psi}_{t}\}$ in $\mathcal{T}$ are set entirely by $H'$. These could be interleaved anywhere within the spectrum of $H_{\rm SM}$, including at an energy density corresponding to infinite temperature. 
The other states at the same energy density in $\overline{\mathcal{T}}$ are however expected to reproduce thermal expectation values, as $H_{A}$ (and thus $H_\mathrm{SM}$) should be non-integrable for generic Hermitian $h_i$. 
For example, the expectation value of the local projector $P_i$ in any of these states should be proportional to $\textrm{tr}(P_i)$. The same expectation value in a scarred eigenstate is exactly zero, at odds with the ETH. 

It is useful to briefly discuss the embedded states in relation to possible symmetries of $H_{\rm SM}$. 
The ETH holds only within individual symmetry sectors.
SM presented their embedded states  as counterexamples to the ETH because $H_{\rm SM}$ generically does not have any obvious local symmetries, and so the embedded target states need not live in a separate symmetry sector from the rest of spectrum. Crucially, the $P_i$'s do not generically form a mutually commuting set and hence are not a set of local symmetries for the Hamiltonian~\footnote{However, when the projectors do commute, the disconnected sectors can be interpreted as being distinguished by symmetry~\cite{Mondaini18}.}.

An alternative perspective is to think of $\mathcal T$ and $\overline{\mathcal T}$ as \emph{dynamically} disconnected subspaces which do not couple to each other under the action of $H_{\rm SM}$. In other words, $H_{\rm SM}$ can decomposed as
\begin{equation}
    H_{\rm SM} = H^{\rm scar}_{ \mathcal{T}} \oplus H^{\rm thermal}_{\overline{\mathcal{T}}}.
\end{equation}

Several examples of embedded towers of states with quasiparticles lie within the SM construction. Remarkably, the spin-1 XY bimagnon scar tower also falls into this class, although the model was not explicitly constructed using the SM formalism. The projection operators associated with the embedding are given by
\begin{align}
\begin{split}
    P_r=1&-\frac{3}{4}(S^z_r)^2(S^z_{r+1})^2+\frac{1}{8}\left[(S^+_r)^2(S^{-}_{r+1})^2+\text{H.c.}\right]- \frac{1}{4}S^z_r S^z_{r+1}.
\end{split}
\end{align}
We refer the reader to Appendix C of Ref.~\cite{Schecter19} for a more detailed discussion. 
The SM method has also been used to embed a macroscopic SU(2) spin~\cite{Choi19}, multiple towers of magnon states~\cite{Tang21} and magnon crystals~\cite{McClarty:2020,Kuno:2020}, as well as states with no quasiparticle interpretation, \emph{e.g.} zero modes in gauge theories~\cite{Banerjee21,Biswas22}, topologically ordered states~\cite{Ok:2019,Srivatsa20b,Wildeboer21} and states in geometrically frustrated systems~\cite{Lee:2020, McClarty:2020}. 

\subsection{Hilbert Space Fragmentation}

\begin{figure}[t!]
\includegraphics[width=1.00\columnwidth]{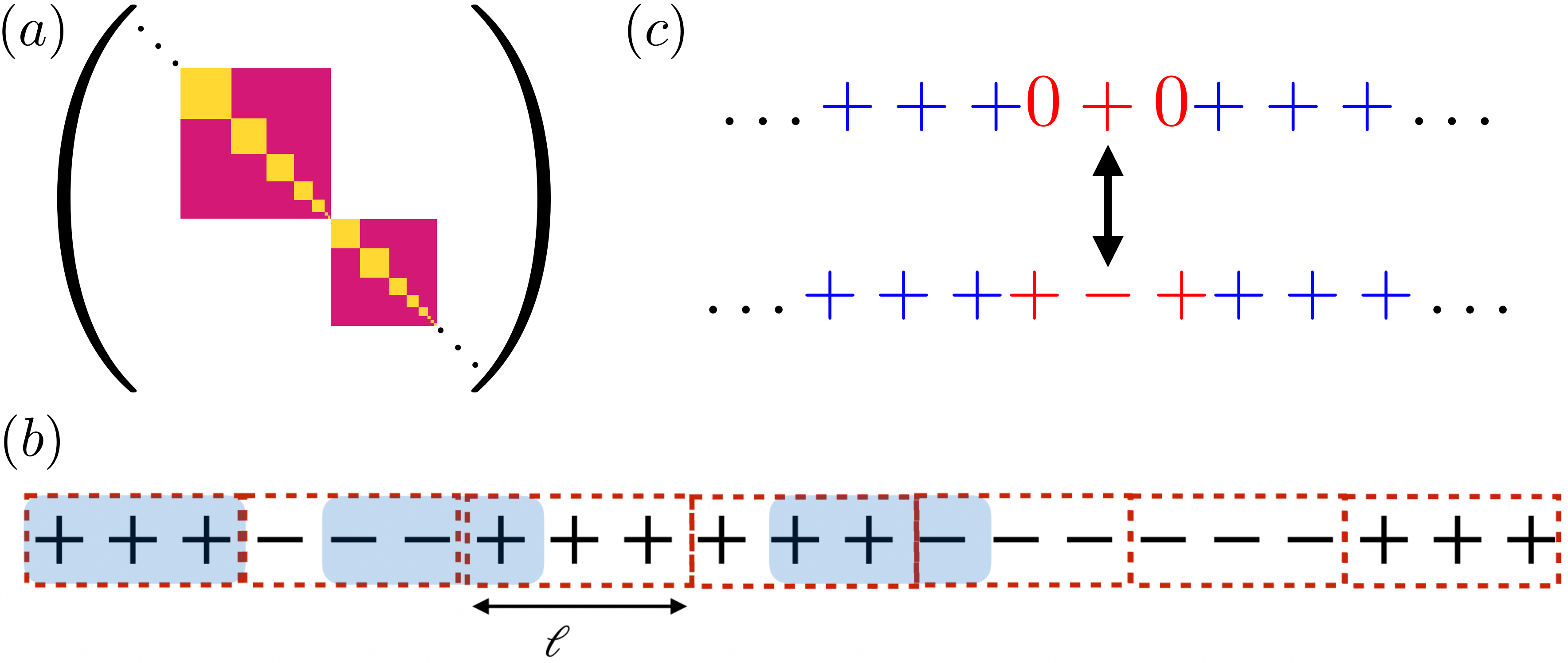}
\caption{
A mechanism for Hilbert space fragmentation (panel (b) adapted from Ref.~\cite{Khemani20}). (a) Schematic of the block structure of the evolution operator in a system with Hilbert-space fragmentation. The pink blocks denote different symmetry sectors, while the yellow blocks denote disconnected Krylov subspaces. (b) Example of a frozen state in a one-dimensional spin-1 chain with U(1) charge and dipole-moment conservation acted upon by local gates with range at most $\ell=3$. Any gate acting on three consecutive sites (\emph{e.g.}, the shaded blue regions) cannot change the state because the local configuration either has extremal charge (as in the leftmost block) or dipole moment (as in the middle and rightmost blocks). (c) Example of a two-dimensional Krylov subspace generated by an active region (red) embedded in a frozen background (blue). The transition indicated by the black arrow is the only one allowed by charge and dipole conservation.
}
\label{fig:fragmentation}
\end{figure}

We now turn to a related but conceptually distinct mechanism for ergodicity breaking, which does not rely on eigenstates or even time-periodicity of the dynamics. This phenomenon is variously described as ``Hilbert space shattering"~\cite{Khemani20}, ``Hilbert space fragmentation"~\cite{Sala20}, or ``Krylov fracture"~\cite{Moudgalya19b} and refers to the existence of an exponentially large number of dynamically disconnected sectors of Hilbert space. This fragmentation of Hilbert space can be viewed as a generalization of the emergence of dynamically disconnected subspaces discussed in the previous section.

The most familiar way to obtain dynamically disconnected sectors is through the existence of conservation laws, with different sectors labeled by the values of conserved quantities. Thermodynamically, it is the sector with the lowest free energy which determines the physical behavior, but if a system is initialized in a different symmetry sector, observables will be determined by the statistical properties of that sector. 

On the other hand, interesting examples of Hilbert space fragmentation involve the existence of dynamically disconnected sectors \emph{even after resolving all microscopic symmetries} --- \emph{i.e.}, states with the same symmetry quantum numbers are still forbidden from mixing under the dynamics~\cite{Khemani20, Sala20} (see Fig.~\ref{fig:fragmentation}(a)). The unitary operator governing the time-evolution thus has a block diagonal structure (in some simple basis), which reflects both the symmetries of the problem, \emph{and} further fragmentation \emph{within} symmetry sectors. Strikingly, for systems with one or a few symmetries, while the number of symmetry sectors generally scales at most polynomially with system size, the number of dynamical sectors can scale \emph{exponentially} with system size. The dimension of the individual sectors can range from one (corresponding to a single frozen state) to exponential in system size.  A further distinction between \emph{strong} and \emph{weak} fragmentation can be made depending on whether the ratio of the dimension of the largest dynamical sector to that of its symmetry sector tends to zero or one in the thermodynamic limit, respectively~\cite{Khemani20, Sala20}.
In the former (strong) case, there is an exponential number of dynamically disconnected sectors, which can lead to nonthermalizing dynamics even from typical initial states. The latter (weak) case is more generic, and predicts thermalization to the appropriate thermodynamic ensemble from most initial states.

Fragmented dynamics is considered scarred because of (i) a strong initial state dependence in the dynamics, ranging from fully localized to typically thermal depending on the dynamical sector in which one starts, and (ii) anomalous or frozen dynamics from simple low-entanglement initial states which never thermalize to the equilibrium state appropriate to the symmetry quantum numbers of the initial state. 

The dynamical sectors are also referred to as Krylov sectors, following standard prescriptions for constructing the Krylov subspace $\mathcal{K}$ of a state $|\psi_0\rangle$ time-evolving under a Hamiltonian $H$:  the state $e^{iHt}|\psi_0\rangle$ lives in the span of all vectors obtained by the repeated action of $H$ on $|\psi_0\rangle$, $\mathcal{K} = {\rm span}\{|\psi_0\rangle, H\ket{\psi_0}, H^2\ket{\psi_0}, \cdots\}$. In the cases of most interest, the states $|\psi_0\rangle$ are simple experimentally preparable states (such as product states), and the dimension of $\mathcal{K}$ is a vanishing fraction of the full Hilbert space dimension, leading to ``Krylov restricted thermalization" within $\mathcal{K}$~\cite{Moudgalya19b}; in certain cases, the dynamics in $\mathcal{K}$ may be integrable even if the model as a whole is not~\cite{Moudgalya19b,Yang20}. Recent works have constructed integrals of motion that label disjoint Krylov subspaces. In systems with Hilbert space fragmentation, there are exponentially many such integrals of motion that need not be sums of local operators~\cite{Rakovszky20,Moudgalya22}.

A transparent manifestation of Hilbert space fragmentation occurs in local systems which conserve both a global U(1) charge and its dipole moment~\cite{Pai19,Khemani20, Sala20}. In a one dimensional spin system, these symmetries take the form $Q = \sum_x S_x^z$ and $P = \sum_x x S_x^z$, where the sum is over all lattice sites. These symmetries are characteristic of ``fracton" topological phases (for reviews of fractons, see Refs.~\cite{Nand_2019,You_2020}), and their combination produces $O(L^3)$ symmetry sectors labeled by $(Q,P)$ quantum numbers in a system of size $L$. In contrast, these symmetries are provably sufficient to produce exponentially many dynamical subsectors, including exponentially many \emph{strictly} frozen product states which live in dynamical subspaces of dimension exactly equal to one. In addition to the frozen states, there is a wide distribution of dimensions for the emergent subsectors, leading to a strong initial state dependence in the dynamics. The fragmentation is robust to the presence of both spatial and temporal randomness as long as the symmetries and locality are maintained.

The existence of exponentially many strictly frozen states is readily verified by the following simple argument~\cite{Khemani20}.  It is easiest to think of random unitary circuit models in which each unitary gate has range $\ell$ and each gate locally conserves $Q, P$. Let us denote by $+$/$-$ the maximum/minimum local charges on a site respectively; these could be the ``top" and ``bottom" states of a qudit of spin $S$ so that $S^z = \pm S$. Now, note that any pattern that alternates between locally ``all plus" and locally ``all minus," with domain walls between ``all-plus" and ``all-minus" regions at least $\ell$ sites apart, must be inert by local charge and dipole moment conservation. Fig.~\ref{fig:fragmentation}(b) shows an example of such a frozen configuration with $S=1$ and $\ell=3$. A straightforward lower bound on the number of such frozen states can be obtained by dividing the $L$-site chain up into blocks of size $\ell$. Since all spins in each block must be either $+$ or $-$, there must be at least $2^{L/\ell}$ such frozen states. This is exponentially large in $L$ for {\it any} finite $\ell$, but such frozen states still constitute a vanishing fraction of the full Hilbert space. For a more precise counting, see, \emph{e.g.}, Ref.~\cite{Khemani20}.

Other sectors with dimension larger than one can be systematically built by embedding dynamically active blocks into sufficiently large frozen backgrounds so as to keep the active region localized to a finite region of space; see Fig.~\ref{fig:fragmentation}(c) for a simple example. 


It is also easy to write down Hamiltonians with charge and dipole conservation, for instance Hamiltonians with coordinated ``pair hopping" terms $\ket{1001} \leftrightarrow \ket{0110}$. Such Hamiltonians arise when considering models of the fractional quantum Hall effect in thin cylinders, \emph{i.e.}~in a quasi-1D limit~\cite{Tao83,Bergholtz05,Moudgalya19,Moudgalya19b}. Such pair hopping terms could also emerge as the leading-order description of systems placed in large electric fields~\cite{Moudgalya19b,Khemani20}. Dipole moment couples directly to electric field, so if the system is placed in a sufficiently large static electric field (or equivalently, a tilted potential) then the Hilbert space will (approximately) split into symmetry sectors labelled by dipole moment (equivalently, center of mass position), with states at different dipole moment having sharply different energies. Equivalently, such systems can be viewed as possessing an emergent prethermal conservation of dipole moment that disappears after an exponentially long timescale~\cite{Abanin_17, Mori_16, Khemani20}. Much recent theoretical and experimental \cite{Guo_2021,Wang_2021,Morong_2021} work has focused on such tilted systems, with claims that there may be a phase transition to a so-called Stark MBL~\cite{Schulz_2019,Refael_2019} phase in which a disorder-free system becomes many-body localized above a critical value of the tilt strength. However, it has been argued that the numerical signatures of Stark MBL may instead be finite-size and -time effects due to apparent Hilbert space fragmentation from long-lived prethermal dipole conservation induced by the tilted potential~\cite{Khemani20}. One can also consider conservation of higher multipole moments in higher dimensions, which have been shown to be associated with slow subdiffusive hydrodynamics~\cite{Gromov20,Grosvenor21,Iaconis21}. 

Analogous physics has been studied in various other contexts including, \emph{e.g.}, interacting fermion models~\cite{Moudgalya19b,Iadecola18,DeTomasi19,Nachtergaele:2020ws}, models with kinetic constraints (including in classical settings)~\cite{Lan18, OlmosLesanovsky, Gopalakrishnan18,Sarang_Automaton1,Hudomal:2020,Zhao:2020,Yang20} and multicomponent Hilbert spaces~\cite{Zhao21,Buca22}, and models with emergent gauge fields such as frustrated magnets and dimer models~\cite{McClarty:2020,dimerfracture,Zhang22c}. The existence of dynamically disconnected sectors and the resulting danger of obtaining thermodynamically incorrect results by getting trapped in an atypical sector has also been appreciated for a long time by the numerical simulation community~\cite{dimerfracture,Moessner_rvb}---this is the flipside of using Hilbert space fragmentation to obtain non-thermal behavior. 
This goes along with autonomously interesting investigations of the sector structure of such models, see \emph{e.g.}~Ref.~\cite{Cepas_21}. The possibility of obtaining disorder-free glassy dynamics 
has also been discussed in this context, see \emph{e.g.}~Ref.~\cite{Yin_2001,Garrahan02,Garrahan07,VanHorssen15,Lan18,Pancotti20,Charbonneau:2017ul}. 
However, we note although dynamics in these various other settings can be described by the existence of dynamically disconnected subsectors, there is no understanding of general (\textit{i.e.}, constructive) conditions that lead to Hilbert space fracture in these models and no principled way to examine the stability of fracturing in these models to the addition of perturbations or noise. For example, the simplest kinetically constrained models comprise spin 1/2 systems in which the spin on a site can flip if certain conditions are obeyed by its neighbors, for instance if both neighbors are down. However, there is no unique or natural way to deform such models, for example, to include the effect of further neighbor spins. Likewise, the dynamics in quantum dimer models come from certain “flippable” plaquettes which are lattice dependent. While allowing for longer flippable loops decreases fracture~\cite{dimerfracture,Moessner_rvb}, there are no general results on how the number of disconnected sectors scales with such perturbations. In contrast, the dipole-conserving models discussed here provide a robust symmetry-based constructive mechanism for Hilbert space fracture.

\section{Experiments and the Puzzles They Inspire} 
\label{sec:Experiments}

Much of the present interest in quantum many-body scars originates from an experiment reported in Ref.~\cite{Bernien17}, which observed unexpected coherent oscillations in a system of interacting Rydberg atoms when the array was prepared in a particular initial state. (Further systematic experimental analysis of this phenomenon was undertaken in Ref.~\cite{Bluvstein21}.) 
In this section, we summarize this experiment and the theoretical attempts to understand it, making connections and drawing contrasts with the quasiparticle-centric point of view adopted in this review. We close this section with a summary of other experiments that have been performed in this domain. Throughout, we emphasize the remaining theoretical puzzles.

\subsection{Rydberg Atom Experiment and the PXP Model}

In Ref.~\cite{Bernien17}, a system of rubidium atoms trapped in optical tweezers was used to perform an analog quantum simulation of the one-dimensional Ising model in transverse and longitudinal fields. The array of atoms is driven in such a way that each atom can be either in its ground state or in a highly excited Rydberg state. The resulting system can be viewed as an array of coupled two-level systems, described by Pauli operators $Z_i$ and $X_i$, with Hamiltonian
\begin{align}
\label{eq:Ising}
    H = \sum_{i<j}V_{ij}\, n_i n_j - \Delta \sum_i n_i + \frac{\Omega}{2}\sum_i X_i.
\end{align}
Here, $n_i = (1+Z_i)/2$ is the occupation of the Rydberg state of the $i$th atom and $X_i$ couples the $i$th atom's ground and Rydberg states. The parameter $\Omega$ is the Rabi frequency of the driving laser, and $\Delta$ is the detuning of the driving laser from the Rydberg state. The coupling $V_{ij}>0$ originates from van der Waals interactions between excited atoms, and falls off as $R^{-6}_{ij}$ with the distance $R_{ij}$ between atoms $i$ and $j$. When the nearest neighbor interaction strength $V\equiv V_{i,i+1}$ is much stronger than both $V_{i,i+2}$ and $\Omega$, the system enters the so-called ``Rydberg blockade" regime in which nearest-neighbor atoms cannot simultaneously be excited~\cite{Jaksch00}. The leading order effective Hamiltonian in this regime can be written as
\begin{align}
\label{eq:PXP}
    H_{\rm eff} = \sum_i \left(\frac{\Omega}{2}\, P_{i-1}X_iP_{i+1}-\Delta\, n_i\right),
\end{align}
where the projector $P_i=(1-Z_i)/2$ ensures that the Rydberg-blockade constraint is satisfied, and where longer-range interaction terms $V_{ij}$ with $|i-j|>1$ have been neglected. This effective model has a history dating at least as far back as Ref.~\cite{Fendley04}, and, when $\Delta=0$, is known as the PXP model. The model is also relevant to U(1) lattice gauge theory in its quantum link~\cite{Surace19} and dimer model~\cite{Moessner01,Laumann12,Chen17} formulations, and to the fractional quantum Hall effect at $\nu=1/3$ in the thin-torus limit~\cite{Moudgalya19}. 

{As an aside, the form of the model in Eq.~\eqref{eq:PXP} reflects one of the central themes in the physics of (especially topological) correlated quantum matter. Demanding that the largest term in a Hamiltonian be satisfied requires the identification of, and projection onto, the resulting ``physical" subspace. This is the case here with the Rydberg blockade, and appears analogously in the avoidance of double occupancy in high-temperature superconductors via a Gutzwiller projection \cite{Gutzwiller_63,Anderson_87}, in enforcing the ice rules in spin ice yielding an emergent magnetostatics \cite{Castelnovo_12}, or in minimizing kinetic energy in quantum Hall physics by projecting onto a partially filled Landau level \cite{Girvin_99}.}

The experiment in Ref.~\cite{Bernien17} prepared a N\'eel initial state, which can be expressed in the Rydberg occupation basis as $\ket{\mathbb Z_2}=\ket{101\dots}$, and evolved it with the Hamiltonian in Eq.~\eqref{eq:Ising} with $\Delta=0$ in the Rydberg blockade regime. When $\Delta=0$, the average energy $\braket{\mathbb Z_2|H_{\rm eff}|\mathbb Z_2}$ lies in the middle of the spectrum of the nonintegrable model $H_{\rm eff}$---\textit{i.e.}, $\ket{\mathbb Z_2}$ is effectively an infinite-temperature state in the Rydberg-blockaded subspace. Thus, the expectation based on the ETH is that the dynamics of this initial state should rapidly decohere and local observables should thermalize to their infinite temperature expectation value; nevertheless, coherent oscillations of the occupation numbers $n_i$ were observed about a non-thermal value. Concurrent theoretical work reported in Ref.~\cite{Bernien17} found that the entanglement entropy following this quench from the N\'eel state grows much more slowly than for other product states and also exhibits oscillations at the same frequency. 

Expanding the N\'eel state in the eigenbasis of Eq.~\eqref{eq:PXP} shows that the unexpected oscillations can be traced back to a set of $(L+1)$ special low-entanglement eigenstates in the spectrum of the PXP model for a chain of $L$ sites. This set of eigenstates spans the entire many-body bandwidth of the PXP model in the sector with no nearest-neighbor Rydberg excitations, and includes both the ground state and the highest-energy state. These eigenstates, which were termed quantum many-body scar states in Ref.~\cite{Turner18a}, have anomalously high overlap with the $\ket{\mathbb Z_2}$ state and are approximately equally spaced in energy, leading to coherent oscillatory dynamics of a variety of quantities and periodic near-revivals in the Loschmidt echo $\mathcal L(t)$ [see Eq.~\eqref{eq:S1XYLt}]. Followup work in Ref.~\cite{Turner18b} found that the scar states also have anomalously low entanglement relative to other eigenstates of the PXP model at the same energy density, with half-chain entanglement entropy scaling $\sim\ln L$. 

\subsection{Attempts to Understand Scars in the PXP Model}

While the existence of low-entanglement scarred eigenstates in the PXP model provides a \emph{descriptive} explanation of the experimental observations of Ref.~\cite{Bernien17}, a complete understanding of \emph{why} this model contains scars is still lacking. We now describe various attempts to understand the origin of the scarred eigenstates in the PXP model. 

One line of work originated by the authors of Refs.~\cite{Turner18a,Turner18b}, known as the forward scattering approximation (FSA), constructs the scar states using a Krylov subspace approach. In this approach, the PXP Hamiltonian, $H_{\rm PXP}=\sum_iP_{i-1}X_iP_{i+1}$, is broken up as
\begin{subequations}
\begin{align}
    H_\mathrm{PXP}=H^+ + H^-,
\end{align}
where
\begin{align}
    H^+=\sum_{i\text{ odd}} P_{i-1}\sigma^-_iP_{i+1}+\sum_{i\text{ even}} P_{i-1}\sigma^+_iP_{i+1}
\end{align}
\end{subequations}
and $H^+ = (H^{-})^\dagger$. The operators $H^{\pm}$ act as raising and lowering operators for the staggered magnetization $Z_\pi = \frac{1}{2}\sum_i(-1)^{i}Z_i$, i.e.
\begin{align}
\label{eq:PXPSU2}
    [Z_\pi,H^\pm]=\pm H^{\pm}.
\end{align}
The N\'eel state $\ket{\mathbb Z_2}$ is the minimum-eigenvalue eigenstate of $Z_\pi$. For a chain of $L$ sites, one can therefore define a Krylov subspace $\mathcal K = \{\ket{0},\ket{1},\dots,\ket{L}\}$ spanned by $L+1$ states $\ket{k}\propto(H^+)^k\ket{\mathbb Z_2}$, so that $\ket{0}=\ket{\mathbb Z_2}$ and $\ket{L}=\ket{\mathbb Z'_2}=\ket{010\dots}$. Approximations to the scar states are then obtained by diagonalizing the projection of $H_\mathrm{PXP}$ into the subspace $\mathcal K$, which is a matrix of dimension $L+1$. The FSA thus allows one to obtain approximations to the scarred eigenstates of the PXP model at system sizes much larger than those accessible to exact diagonalization and, as shown in Ref.~\cite{Turner18b}, produces eigenstates with entanglement entropy scaling as $\ln L$. More recently, Ref.~\cite{Desaules21} found a nonlocal interacting model for which the FSA becomes exact. However, this model is not perturbatively connected to the PXP model.

Eq.~\eqref{eq:PXPSU2} is suggestive of an SU(2) algebra generated by $H^\pm$ and $Z_\pi$, and thus of an SGA. However, as noted in Ref.~\cite{Choi19}, this algebra is not closed, since
\begin{align}
\label{eq:Hpm}
    [H^+,H^-]=\frac{1}{2}Z_\pi+\frac{1}{4}O_{ZZZ},
\end{align}
where we have assumed periodic boundary conditions and defined $O_{ZZZ}=\sum_i(-1)^iZ_{i-1}Z_iZ_{i+1}$. We revisit the SGA connection below.

Subsequent work in Ref.~\cite{Turner20} unified the FSA approach with parallel works that described the oscillatory scarred dynamics in the PXP model using the time-dependent variational principle (TDVP)~\cite{Ho19,Michailidis20,Michailidis20revivals}. The TDVP is a semiclassical approach that projects the quantum dynamics of a system into a variational manifold of matrix product states~\cite{Haegeman11}, leading to a set of classical equations of motion that can exhibit phenomena familiar from the theory of classical chaos. In this way, the FSA can be brought into contact with the theory of quantum scars in single-particle quantum mechanical systems arising as quantizations of classical dynamical systems~\cite{Heller84}. However, while the TDVP approach captures the scarred dynamics in the space of variational matrix-product states with low bond dimension (and hence low entanglement entropy), the periodic orbits may disappear on increasing the bond dimension~\cite{Michailidis20revivals}. More work is thus required to sharpen the connection to single-particle scars.  

A parallel strand of works have taken inspiration from the tower-like structure of the scar states, as evidenced by their nearly constant energy spacing and logarithmically scaling entanglement entropy, to attempt to formulate a quasiparticle description of these states. 

One manner in which a quasiparticle description of the scar states can arise is through proximity to an integrable model, where \textit{every} eigenstate can be described in terms of quasiparticles~\cite{Franchini:2017vz}. The proximity to integrability further guarantees that the dynamical response of the integrable model survives for a parametrically long time~\cite{Berges:2004,Kinoshita:2006ve,Gring:2012ly,Neyenhuis:2017wi,Eckstein:2009wv,Eisert:2015aa,Vasseur:2016ub}. The connection between the PXP model and integrability was first pointed out in Ref.~\cite{Khemani19},  which found that adding   
\begin{align}
\label{eq:PXPZ}
    H_\mathrm{XZ}=h_\mathrm{XZ}\sum_i P_{i-1}X_iP_{i+1}(Z_{i-2}+Z_{i+2})
\end{align}
to $H_\mathrm{PXP}$ with a small coefficient $h_\mathrm{XZ}\simeq -0.02$ both strengthened the revivals and pushed the model closer towards integrability (as measured by various diagnostics such as Poisson statistics in energy-level spacings and fluctuations of eigenstate expectation values). Intriguingly, $H_\mathrm{XZ}$ and its longer-range generalizations also enhance the accuracy of the FSA~\cite{Choi19}---albeit with a different value of $h_{\mathrm XZ}$---and have a simple physical origin in complete bipartite graphs~\cite{Windt21}.  However, whether there exists an exact integrable point in the vicinity of $H_\mathrm{PXP} + H_\mathrm{XZ}$ remains unknown.

Another possibility is that the PXP model is proximate to a nonintegrable model with a \textit{single} infinitely long-lived quasiparticle species atop a vacuum state, similar to the case of the spin-1 XY model discussed in Sec.~\ref{sec:S1XY}. (Of course, this tower may still be obtained by adding an appropriate $H_A$ to a parent integrable/high-symmetry model as discussed in Section~\ref{sec:QPConstructive}.) 
Ref.~\cite{Iadecola19} constructed an approximate tower of momentum-$\pi$ magnon states atop the ground state of the PXP model and showed that the scar states found in Ref.~\cite{Turner18a} have high overlap with this tower. The ground state, which we call $\ket{E_0}$, is known to be paramagnetic and has an efficient variational representation given in Ref.~\cite{Ovchinnikov03}. The lowest-lying excitation atop this ground state was shown numerically in Ref.~\cite{Iadecola19} to have high overlap with the state $S^+_\pi\ket{E_0}$, where
\begin{align}
\label{eq:PXPSp}
    S^\pm_\pi=Z_\pi\mp i\alpha\, Y_\pi,
\end{align}
where $Z_\pi$ is the staggered magnetization, $Y_\pi=\frac{1}{2}\sum_i (-1)^i P_{i-1}Y_iP_{i+1}$, and $\alpha \approx 2$ is a constant determined by numerical optimization. This excitation is naturally interpreted as a magnon with momentum $\pi$. Similar excitations can be created atop the highest excited state of the model, $\ket{E_{\rm max}}$, via $S^-_\pi\ket{E_{\rm max}}$. This observation motivates the definition of a variational subspace spanned by $\pi$-magnon states,
\begin{align}
\mathcal V_\pi = \text{span}\{(S^+_\pi)^n\ket{E_0},(S^-_\pi)^m\ket{E_{\rm max}}\}^{L/2}_{n,m=0},
\end{align}
in which the tower of scar states have substantial weight at finite system sizes. Moreover, enlarging the variational basis with states in which magnons have nontrivial relative momenta, e.g. $(S^+_\pi)^{n-2}S^+_{\pi-\delta k}S^+_{\pi+\delta k}\ket{E_{0}}$, systematically increases the weight of the scar states in the variational magnon subspace. 

Let us now revisit the connection to SGAs. Setting $\alpha=2$ in Eq.~\eqref{eq:PXPSp}, one finds that~\cite{Iadecola19}
\begin{align}
\label{eq:PXPSU22}
\begin{split}
    [S^+_\pi,S^-_{\pi}]&=2\, H_{\rm PXP}\\
    [H_{\rm PXP},S^{\pm}_\pi]&=\pm S^\pm_{\pi}\pm O_{ZZZ},
\end{split}
\end{align}
where $O_{ZZZ}$ is defined below Eq.~\eqref{eq:Hpm} Up to the extra $O_{ZZZ}$ term in the second line, this is an SU(2) algebra. 
Eq.~\eqref{eq:PXPSU22} can be viewed as analogous to Eqs.~\eqref{eq:PXPSU2} and \eqref{eq:Hpm}, except that the roles of $Z_\pi$ and $H_{\rm PXP}$ have been interchanged. This point of view suggests a conceptual link between the FSA and the quasiparticle picture of Ref.~\cite{Iadecola19}. 

One consequence of the algebra in Eq.~\eqref{eq:PXPSU22} is the prediction of off-diagonal long-range order in the scar states (see the discussion around Eq.~\eqref{eq:J+J-}). Such eigenstate order would result in coherent oscillations of the two-time correlator $\mathcal C_Y(t)=\braket{\mathbb Z_2|Y_\pi(t)Y_\pi(0)|\mathbb Z_2}/L^2$ in the thermodynamic limit~\cite{Iadecola19,Chen22}.

Alternative quasiparticle descriptions of the scar states in the PXP model have also been proposed. Ref.~\cite{Pan22} constructs a quasiparticle tower of states atop the ground state, similar to Ref.~\cite{Iadecola19}, except it starts from a variational approximation to the ground state developed in Ref.~\cite{Lesanovsky12a}. In Ref.~\cite{Lin18}, several exact matrix-product-state eigenstates of the PXP model were found near the middle of the energy spectrum, including at zero energy where the model exhibits an exponentially large degeneracy~\cite{Turner18a,Schecter18}. Ref.~\cite{Lin18} constructed variational approximations to scar states near the middle of the spectrum by acting on the exact zero-energy matrix-product eigenstates with local operators determined by numerical optimization. Note that this quasiparticle picture predicts that the scar states near the middle of the spectrum ultimately have area-law entanglement, since acting with a local operator on a finite-bond-dimension matrix product state can only increase the entanglement entropy by a finite amount in the thermodynamic limit. This would contradict the numerical finding of Ref.~\cite{Turner18b} that the scar states have entanglement entropy $\sim\ln L$.

More recently, another quasiparticle picture was proposed in Ref.~\cite{Omiya22}. In that work, the PXP model is converted to an effective spin-1 chain using a mapping proposed in Ref.~\cite{Lin18}. A tower of states is then constructed by acting on the fully polarized state of the effective spin-1 chain with an appropriate spin-1 lowering operator, followed by a projection operator that enforces the Rydberg blockade constraint. This tower of states has the appealing property that the N\'eel state $\ket{\mathbb Z_2}$ can be written exactly as a superposition of the states in the tower, similar to the case of the spin-1 XY model where the initial state $\ket{\phi_0}$ (Eq.~\eqref{eq:Initial state}) can be written as a superposition of scar states. This work found numerically that the scar states of the PXP model have high overlap with this tower. Intriguingly, it also constructs a non-Hermitian extension of the PXP model for which the states in this tower are exact eigenstates.

One common theme that has arisen throughout the works described in this section is the hypothesis that the PXP model is proximate to a model with exact scars. The perturbation, Eq.~\eqref{eq:PXPZ}, proposed in Ref.~\cite{Khemani19} and its longer range variants considered in Refs.~\cite{Choi19,Bull19b} enhance revivals that are characteristic of scarred dynamics. These terms also enhance the accuracy of Ref.~\cite{Turner18a}'s FSA, and of Ref.~\cite{Iadecola19}'s quasiparticle tower picture. Furthermore, Ref.~\cite{Omiya22} found that terms of this form naturally emerge when trying to make their non-Hermitian model Hermitian. Reconciling these various perspectives where possible and determining the relationship (if any) to the model with an exact FSA proposed in Ref.~\cite{Desaules21} remains an interesting research direction.

\subsection{Other Experiments}
A variety of other experiments have detected signatures of many-body scarring. First, several recent experiments have realized the PXP model in other platforms. For example, Ref.~\cite{Su22} used a Bose-Hubbard system in a tilted optical lattice to simulate the PXP model and measured entanglement and fidelity dynamics. Ref.~\cite{Chen22} used superconducting-qubit processors from IBM to realize the nearest-neighbor limit of the mixed-field Ising model, Eq.~\eqref{eq:Ising}, and probe the dynamics in the scarred regime, including measurements of the nontrivial correlator $\mathcal C_Y(t)$. Another extension of the work performed in Ref.~\cite{Bernien17} is Ref.~\cite{Bluvstein21}, which studied a periodically driven version of the PXP model and found intriguing subharmonic features in the dynamics of the N\'eel state.

Scars have also been realized in experimental platforms simulating other models. For example, in Ref.~\cite{Zhang22} scarred dynamics were observed in a superconducting-qubit processor performing analog quantum simulation of a spin-1/2 XY model. In this work, the integrability of the spin-1/2 XY model is broken by random longer-range cross-coupling terms arising due to experimental imperfections, and scarred dynamics are observed from certain special initial states. Ref.~\cite{Wil:2021vr} instead creates long-lived, highly excited but non-thermal states in a bosonic one-dimensional quantum gas of dysprosium using a topological pumping mechanism. Starting from the ground state, the authors cyclically vary the contact interaction strength; as the Hamiltonian of the gas is close to an integrable one (the Lieb–Liniger model~\cite{Giamarchi:2003oq,Cazalilla:2011nx}), this process maps one eigenstate to another higher up in energy instead of heating up the gas.

\section{Conclusion and Outlook}

In this review, we have considered the phenomenon of quantum many-body scarring through the lens of quasiparticles. We began by considering a relatively simple model of quantum many-body scars, the spin-1 XY model, where the existence of rare, nonthermal eigenstates can be shown analytically. The nonthermal eigenstates consist of a finite density of quasiparticles, known as bimagnons, with momentum $\pi$, whose scattering and dissociation are prevented by a destructive interference mechanism. We then moved on to consider more general settings for the realization of such towers of quasiparticle states, which can be formally understood in terms of spectrum generating algebras and associated symmetry principles. We also reviewed a few constructive approaches to realize models with towers of many-body scar states. Subsequently, we considered a range of other contexts in which the nomenclature of scarring has been invoked, focusing in particular on projector Hamiltonians and the more extreme phenomenon of Hilbert-space fragmentation. While these mechanisms for realizing nonthermal quantum many-body states have been associated with the phenomenon of scarring, they need not be associated with a quasiparticle description. Finally, we discussed experimental realizations of many-body scars, focusing in particular on the Rydberg experiment of Ref.~\cite{Bernien17} and on the theoretical puzzles that it presents.  Quasiparticles again appear in various guises in theoretical works attempting to explain these experimental results, but are not common to all theoretical descriptions that have been offered.

A variety of open questions remain. One of these concerns the stability of scarred many-body eigenstates to perturbations, which is crucially important both theoretically and in efforts to realize many-body scars in experiments.
One way to phrase this question precisely is to ask what is the timescale for an exact scarred eigenstate $\ket{\psi_0}$ of a nonintegrable Hamiltonian $H_0$ to thermalize under dynamics generated by $H_0+\lambda V$, where $V$ is some local perturbation and $\lambda$ is a small parameter.
Ref.~\cite{Lin19} presented an analysis of this problem on general grounds using Lieb-Robinson bounds and found that the thermalization time can be lower-bounded by $t_*\sim O(\lambda^{-1/(1+d)})$, where $d$ is the spatial dimension of the system.
This lower bound is extremely general and is likely not tight in many cases of interest. Whether tighter lower bounds can be obtained, and understanding the mechanisms by which the thermalization time can exceed this bound in specific cases, are important questions for future work.

Despite the large volume of theoretical work on the subject, another question that remains is to understand the origin of many-body scarring in the PXP model that is relevant to the original Rydberg experiment. Specifically, it would be desirable to understand whether there is a model, integrable or otherwise, in which the scarred eigenstates can be written down exactly. Even if such a model is found, however, there are other unusual features of the PXP model---for example, there are other spin-density-wave-like states, such as $\ket{\mathbb Z_3}=\ket{100100\dots}$, that exhibit approximate revivals~\cite{Turner18b}, and other eigenstates that have anomalously high overlap with the N\'eel state $\ket{\mathbb Z_2}$~\cite{MondragonShem21}. Understanding these features of the model remains an outstanding challenge.

While this review has focused on many-body scars in Hamiltonian systems, they can also occur in other contexts. For example, many-body scars can appear in periodically driven systems, which do not conserve energy and are therefore expected to heat up to infinite temperature at late times~\cite{Lazarides14,D'Alessio14}. There have been various studies of how such a heat death may be avoided, at least on long timescales \cite{Mori_16,Abanin_17,Haldar_21}. A variety of works, including the experiment in Ref.~\cite{Bluvstein21}, have studied periodically driven variants of the PXP model~\cite{Mukherjee19,Mizuta:2020,Sugiura:2021,Maskara21,Iadecola20b,Wilkinson20,Rozon21} and found both exact and approximate scarring. Notably, for a particular choice of periodic driving, the PXP model can be fine-tuned to integrability~\cite{Iadecola20b,Wilkinson20,Rozon21}. Refs.~\cite{Bluvstein21} and \cite{Maskara21} propose an intriguing connection between scarring and discrete time crystals~\cite{Khemani16,Sacha17,Khemani19b,Else20} that deserves further exploration. Quantum scarring in the single-particle context can also occur in non-Hermitian systems~\cite{Wisniacki08}, which motivates considering the possibility of many-body scars in this setting~\cite{Buca19}. While there has been some initial progress in this direction~\cite{Pakrouski21,Chen22,Omiya22}, it would be interesting to consider more broadly the possibility of many-body scarring in open quantum systems. The notion of inverse scars, where thermal states are embedded in a non-thermal spectrum~\cite{Srivatsa_20}, may also be an interesting subject. 

Finally, another important question concerns the range of systems in which many-body scars can be realized experimentally. So far, the vast majority of experimental efforts have focused on realizing the PXP model in different contexts. Only recently have experiments turned their attention to other models; for example, Ref.~\cite{Jepsen:2021} realized long-lived spin-helical states in a lithium atom system simulating anisotropic Heisenberg chains, and Ref.~\cite{Wil:2021vr} obtained scars in a dysprosium gas via topological pumping. However, whether scars stemming from other models and mechanisms could be observed in other experimental platforms is largely unexplored. The question of experimental realization is closely related to the exploration of potential applications for systems realizing quantum many-body scars. 
For example, Refs.~\cite{Dooley21,Desaules21b} show that systems with many-body scars can be useful in quantum sensing and metrology.
Finding other potential applications of this fascinating phenomenon in quantum information science and technology is another worthy direction for future exploration.

\section*{DISCLOSURE STATEMENT}
The authors are not aware of any affiliations, memberships, funding, or financial holdings that
might reasonably be perceived as affecting the objectivity of this review. 

\section*{ACKNOWLEDGMENTS}
We are grateful to Nick O'Dea for preparing Fig.~\ref{fig:S1XY}(b) and for previous collaboration, and to Andrei Bernevig and Maksym Serbyn for providing helpful feedback on the manuscript. 
We also thank our collaborators on previous works on scars and related topics in quantum dynamics, including
Benjamin Burdick, Fiona Burnell, I-Chi Chen, Alexey Gorshkov,
Christopher Langlett, Chris Laumann, Cheng-Ju Lin, Fangli Liu, Olexei Motrunich, Anne Nielsen, Peter Orth,  Michael Schecter, N. Srivatsa, Long-Hin Tang, Julia Wildeboer, Shenglong Xu, Zhicheng Yang, and  Yongxin Yao.
We acknowledge valuable discussions with Jean-Yves Desaules, Wen Wei Ho, Timothy Hsieh, Ana Hudomal, Benjamin Lev, Sanjay Moudgalya, Zlatko Papi\'c, Abhinav Prem and Nicolas Regnault. 

This work was supported in part by the National Science Foundation through awards DMR-1752759 (A.C.) and DMR-2143635 (T.I.), by the US Department of Energy, Office of Science, Basic Energy Sciences, under Early Career Award No. DE-SC0021111 (V.K.), and by the Deutsche Forschungsgemeinschaft under cluster of excellence ct.qmat EXC 2147 project-id 390858490 (R.M.). V.K. also acknowledges support from the Sloan Foundation through a Sloan Research Fellowship and the Packard Foundation through a Packard Fellowship.
%

\bibliographystyle{unsrt11}

\bibliography{refs.bib}

\end{document}